\documentstyle[aps,prl,preprint,floats,epsfig]{revtex}

\begin{document}


\preprint{\tighten\vbox{\hbox{\hfil CLNS {\it 02/1781}}
                 \hbox{\hfil CLEO {\it 02-4}}
                \hbox{\hfil{\it Phys. Rev. D Rapid Communications}}
                 \hbox{\hfil to be published}
}}

\tighten

\title{\large Search for Lepton-Flavor-Violating Decays of $B$ Mesons}

\author{CLEO Collaboration}

\date{9 May 2002}
\maketitle

\begin{abstract}
We have searched a sample of 9.6 million $B \bar B$ events for the
lepton-flavor-violating decays $B \rightarrow h e^\pm \mu^\mp$,
$B^+ \rightarrow h^- e^+ e^+$, $B^+ \rightarrow h^- e^+ \mu^+$, and
$B^+ \rightarrow h^- \mu^+ \mu^+$, where $h$ is $\pi$, $K$, $\rho$, and
$K^*(892)$, a total of sixteen modes.  We find no evidence for these decays,
and place 90\% confidence level upper limits on their branching fractions that
range from  1.0 to $8.3\times 10^{-6}$.
\end{abstract}

\newpage

\begin{center}
K.~W.~Edwards,$^{1}$
R.~Ammar,$^{2}$ D.~Besson,$^{2}$ X.~Zhao,$^{2}$
S.~Anderson,$^{3}$ V.~V.~Frolov,$^{3}$ Y.~Kubota,$^{3}$
S.~J.~Lee,$^{3}$ S.~Z.~Li,$^{3}$ R.~Poling,$^{3}$
% ...
\end{center}
 
\small
\begin{center}
$^{1}${Carleton University, Ottawa, Ontario, Canada K1S 5B6 \\
and the Institute of Particle Physics, Canada M5S 1A7}\\
$^{2}${University of Kansas, Lawrence, Kansas 66045}\\
$^{3}${University of Minnesota, Minneapolis, Minnesota 55455}\\
% ...
$^{11}${Southern Methodist University, Dallas, Texas 75275}\\
% ...
$^{13}${University of Texas - Pan American, Edinburg, Texas 78539}\\
$^{14}${Vanderbilt University, Nashville, Tennessee 37235}\\
% ...
$^{20}${Cornell University, Ithaca, New York 14853}\\
% ...
\end{center}


    The Standard Model predicts that the branching fractions for the decays
$b \rightarrow s e^+ e^-$ and $b \rightarrow s \mu^+ \mu^-$ will be small but
non-zero, of order $10^{-5}$.  We have previously conducted searches for those
inclusive decays\cite{Skwarnicki} and also for the exclusive decays
$B \rightarrow K \ell^+ \ell^-$ and $B \rightarrow K^*(892) \ell^+ \ell^-$
\cite{CLEO-old,KLL} that would result from the quark-level processes.
Others\cite{CDF,BaBar,Belle} have also searched for the exclusive decays.  Upper
limits are now close to the Standard Model predictions, and there is evidence
for $B \rightarrow K \ell^+ \ell^-$\cite{Belle}.

    In contrast, the Standard Model predicts that the topologically similar, but
lepton-flavor-violating decays $b \rightarrow s e^\pm \mu^\mp$ and
$b \rightarrow d e^\pm \mu^\mp$ vanish identically, as do the decays
$B^+ \rightarrow X_s^- \ell^+ \ell^+$ and $B^+ \rightarrow X_d^- \ell^+ \ell^+$.
These decays are predicted to occur in many theories ``beyond the Standard
Model'', for example multi-Higgs extensions\cite{Sher-Yuan}, theories with
leptoquarks\cite{Davidson}, and theories with Majorana neutrinos\cite{Zuber}.
The recent evidence\cite{neut-mix}
that neutrinos mix, and therefore have mass, while not
leading to predictions of observable rates for lepton-flavor-violating decays
involving charged leptons, nonetheless heightens interest in them, as does the
recent claim\cite{double-beta} of neutrinoless double beta decay.

While the underlying physics of lepton-flavor-violating decays is very
different from that of those decays mentioned in the first paragraph, 
the experimental approach in searching for them is quite similar.
We have therefore used the techniques described in
Ref.~\cite{KLL} to search for\footnote{Throughout this
article, the symbol $K^*$ means $K^*(892)$.} $ B \rightarrow K e^\pm \mu^\mp$,
$ B \rightarrow K^* e^\pm \mu^\mp$,
$ B \rightarrow \pi e^\pm \mu^\mp$, and
$ B \rightarrow \rho e^\pm \mu^\mp$, and also for $B^+ \rightarrow h^- e^+ e^+$,
$h^- e^+ \mu^+$, and $h^- \mu^+ \mu^+$, where $h^-$ is $K^-$, $K^{*-}$, $\pi^-$,
and $\rho^-$.  We have previously\cite{Skwarnicki}  searched for the inclusive
decay $b \rightarrow s e^\pm \mu^\mp$, obtaining a 90\% confidence level upper
limit
${\cal B}(b \rightarrow s e^+ \mu^-) + {\cal B}(b \rightarrow s e^- \mu^+)
< 2.2 \times 10^{-5}\ $.
The BaBar collaboration has also searched for, and
reported\cite{BaBar} limits on, the related exclusive decays,
${\cal B}(B^+ \rightarrow K^+ e^\pm \mu^\mp) < 0.8 \times 10^{-6}\ $,
${\cal B}(B^0 \rightarrow K^0 e^\pm \mu^\mp) < 4.1 \times 10^{-6}\ $,
${\cal B}(B^+ \rightarrow K^{*+} e^\pm \mu^\mp) < 8.0 \times 10^{-6}\ $, and
${\cal B}(B^0 \rightarrow K^{*0} e^\pm \mu^\mp) < 3.3 \times 10^{-6}\ $.

    The data used in this analysis were taken with the CLEO
detector\cite{detector} at the Cornell Electron Storage Ring (CESR), a symmetric
$e^+ e^-$ collider operating in the $\Upsilon({\rm 4S})$ resonance region. The
data sample consists of 9.2 ${\rm fb}^{-1}$ at the resonance, corresponding to
9.6 million $B \bar B$ events,
and 4.5 ${\rm fb}^{-1}$ at a center-of-mass energy 60 MeV below the resonance.
The sample below the resonance provides information on the background from
continuum processes $e^+ e^- \rightarrow q \bar q,\ q = u,d,s,c$, and was used
as a check on our Monte Carlo simulation of this background.

    Summing over $e^+ \mu^-$ and $e^- \mu^+$, 
we search for $B \rightarrow K e^\pm \mu^\mp$ in both the $K^\pm$ and
$\bar K^0$ modes, and for $B \rightarrow K^* e^\pm \mu^\mp$ in the
$K^{*0} \rightarrow K^+ \pi^-$ and $K^0 \pi^0$ modes and in the
$K^{*\pm} \rightarrow K^\pm \pi^0$ and $K^0 \pi^\pm$ modes, a total of 6
experimentally distinct final states. (Throughout this article, charge 
conjugate modes are implied.)  Similarly, we
search for $B \rightarrow \pi e^\pm \mu^\mp$ in both the $\pi^\pm$ and $\pi^0$
modes, and for $B \rightarrow \rho e^\pm \mu^\mp$ in both the
$\rho^\pm \rightarrow \pi^\pm \pi^0$ and $\rho^0 \rightarrow \pi^+ \pi^-$ modes,
4 distinct final states.  In the like-sign search
$B^+ \rightarrow h^- \ell^+ \ell^+$, we search for five hadronic final states
($h^- = K^-$; $\pi^-$; $K^{*-} \rightarrow K^- \pi^0, \ \ K^0 \pi^-$;  and
$\rho^- \rightarrow \pi^- \pi^0$) for each of $e^+ e^+$, $e^+ \mu^+$, and
$\mu^+ \mu^+$, 15 distinct modes.  The $K^0$ candidates are
detected via the $K^0 \rightarrow K^0_S \rightarrow \pi^+ \pi^-$ decay chain;
$\pi^0$ candidates via $\pi^0 \rightarrow \gamma \gamma$.

    For those decay modes involving a charged kaon, we use specific ionization 
($dE/dx$) and time-of-flight information to identify the kaon, cutting loosely
(3 standard deviations) if those variables deviate from the mean for kaons in
the direction away from the mean for pions, and cutting harder (1.5 to 2.2
standard deviations, depending on mode) if they deviate on the side towards
the pions.

    There are three main sources of background:\footnote{Throughout this
article, the symbols $\psi$ and $\psi^{\prime}$ mean $J/\psi$(1S) and
$\psi$(2S), respectively.} 
$B \rightarrow K^{(*)} \psi^{(\prime)},\ \psi^{(\prime)} \rightarrow \ell^+
\ell^-$, and other $B \rightarrow \psi^{(\prime)} X$ decays; 
$B \bar B$ decays other than $B \rightarrow \psi^{(\prime)} X$, with two
apparent leptons (either real leptons or hadrons misidentified as leptons);
and continuum processes with two apparent leptons.

    In our previous search\cite{KLL}, for $B \rightarrow K^{(*)} e^+ e^-$ and 
$B \rightarrow K^{(*)} \mu^+ \mu^-$, the backgrounds from $\psi$ and
$\psi^\prime$ were severe.  In the searches reported here they are much less
of a problem, appearing only when particles are misidentified.  Examples are
$B^- \rightarrow K^- \psi,\ \ \psi \rightarrow e^+ e^-$, with the $K^-$
misidentified as a $\mu^-$, and the $e^-$ misidentified as a $K^-$;
$B^- \rightarrow K^- \psi,\ \ \psi \rightarrow \mu^+ \mu^-$, with the $K^-$
misidentified as $\mu^-$, and the $\mu^+$ misidentified as $\pi^+$;
$B^- \rightarrow K^- \psi,\ \ \psi \rightarrow e^+ e^-$, with one of the
$e^\pm$ identified as $\mu^\pm$.  To reduce these backgrounds, we required
that a lepton candidate that passes identification criteria both for $e^\pm$
and $\mu^\pm$ {\it only} be considered as an electron candidate.  Also, we
discarded a candidate reconstruction if any oppositely-charged hadron-lepton
pair, if interpreted as a lepton-lepton pair, had a pair mass within 30 MeV of
$\psi$ or $\psi^\prime$ mass, or if the $e^\pm \mu^\mp$ pair, if interpreted
either as $e^+ e^-$ or $\mu^+ \mu^-$, had a pair mass within 50 MeV of $\psi$ or
40 MeV of $\psi^\prime$.  With these requirements, backgrounds from $\psi$ and
$\psi^\prime$ were rendered negligible, less than 0.1 event per decay mode.

    We discriminate between signal events and the remaining two background
sources using an unbinned maximum likelihood method, including four variables in
the likelihood function.  (We select events for consideration by first applying
loose cuts in those variables.)
    To help distinguish between signal and the background from $B \bar B$ 
semileptonic decays, we use the event missing energy, $E_{\rm miss}$, since 
events with leptons from semileptonic $B$ or $D$ decay  contain neutrinos,
and thus will have missing energy.  We apply loose cuts,
$-2.0 < E_{\rm miss} < +2.0$ GeV.
To help distinguish between signal and continuum events, we use a Fisher
discriminant, a linear combination of $R_2$ (the ratio of second and zeroth
Fox-Wolfram moments\cite{Fox-Wolfram} of the event), 
$\cos \theta_{tt}$ (the cosine of the angle between the
thrust axis of the candidate $B$ and the thrust axis of the rest of the event),
$S$ (the sphericity), and $\cos \theta_B$ (the cosine of 
the production angle of the candidate $B$, relative to the beam direction).  
In particular,
${\cal F} = R_2 + 0.117 \vert \cos \theta_{tt} \vert + 0.779 (1 - S)
+ 0.104 \vert \cos \theta_B \vert$, with values ranging from 0.0 to +2.0.
The coefficients of all terms but $R_2$ were determined by the standard Fisher
discriminant procedure\cite{fisher}.  The relative weight given to $R_2$ was
determined visually, from a scatter plot of $R_2$ {\it vs.} the Fisher
discriminant from the other three variables.  This Fisher discriminant is
identical to the one we used in Ref.~\cite{KLL}.  We apply loose cuts,
$0.0 < {\cal F} < 1.08$.
    Our third and fourth variables used in the likelihood function are the
signal-candidate $B$ reconstruction variables conventionally used for decays
from the $\Upsilon(4S)$: beam-constrained mass
$M_{\rm cand} \equiv \sqrt{E^2_{\rm beam} - P^2_{\rm cand}}$ and
$\Delta E \equiv E_{\rm cand} - E_{\rm beam}$.
Our resolution in $M_{\rm cand}$ is 2.5 MeV, and in $\Delta E$, 20 MeV.  We
apply loose cuts, $5.20 < M_{\rm cand} < 5.30$ GeV and
$-0.25 < \Delta E < +0.25$ GeV.

    We thus have a likelihood function that depends on four variables:
$M_{\rm cand}$, $\Delta E$, $E_{\rm miss}$, and ${\cal F}$.  We vary the
branching fraction for the signal and the yields for the two backgrounds, to
maximize the likelihood.  Probability density
functions (PDFs) are obtained from Monte Carlo samples of continuum events,
$B \bar B$ events, and signal events.  For signal events, lacking a compelling
theoretical model, we use 3-body phase space, with final-state radiation as
given by the CERNlib subroutine {\sc Photos}\cite{photos}.

\begin{figure}
\begin{center}
\epsfxsize=7in
\epsfbox{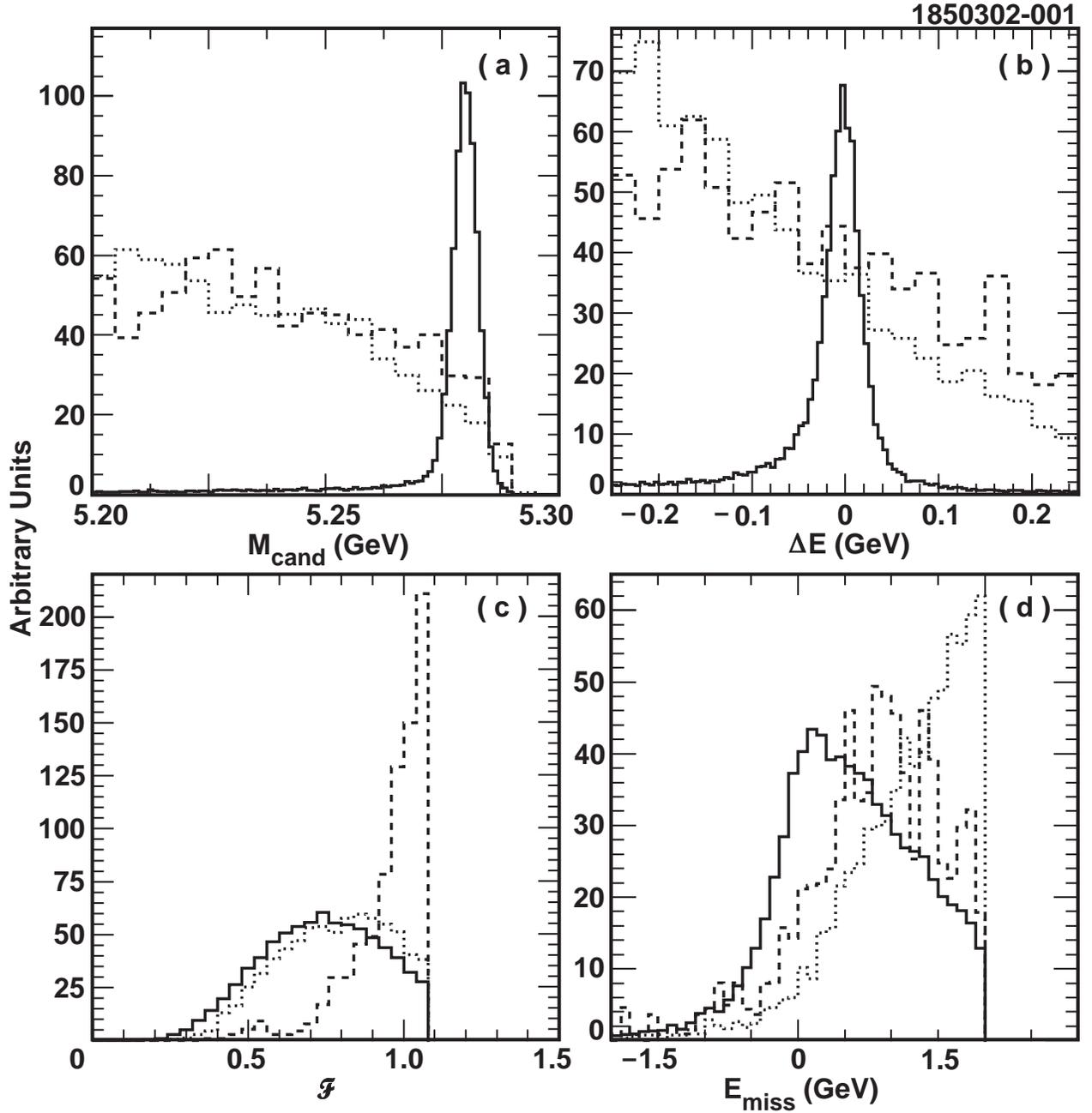}
\hfill \caption{
Distributions in (a) $M_{\rm cand}$, (b) $\Delta E$, (c) ${\cal F}$, and
(d) $E_{\rm miss}$  for Monte Carlo samples of signal events (solid),
$B \bar B$ background events (dotted), and continuum background 
events (dashed), for the search for $B \rightarrow K^{(*)} e^\pm \mu^\mp$.  
The vertical scale is arbitrary.
\label{fig:PDFs}}
\end{center}
\end{figure}

Correlations among the four
variables are weak, both for signal and backgrounds, and we ignore them.
Distributions in the four variables, for signal and the two backgrounds, are
shown for $B \rightarrow K^{(*)} e^\pm \mu^\mp$ in Fig.~\ref{fig:PDFs}.
Distributions for $B^+ \rightarrow h^- \ell^+ \ell^+$ are similar.

    For the decays whose quark-level process is $b \rightarrow s e^\pm \mu^\mp$,
we assume the branching fraction relations
${\cal B}(B^- \rightarrow K^- e^\pm \mu^\mp) =
{\cal B}(\bar B^0 \rightarrow \bar K^0 e^\pm \mu^\mp)$ and
${\cal B}(B^- \rightarrow K^{*-} e^\pm \mu^\mp) =
{\cal B}(\bar B^0 \rightarrow \bar K^{*0} e^\pm \mu^\mp)$, imposing the
equalities as constraints in the maximum likelihood procedure.  Thus our results
here are for the average branching fraction
${\cal B}(B \rightarrow K e^\pm \mu^\mp) \equiv
0.5 ({\cal B}(B^- \rightarrow K^- e^\pm \mu^\mp) +
{\cal B}(\bar B^0 \rightarrow \bar K^0 e^\pm \mu^\mp))$, and similarly with
$K^*$ replacing $K$.
    For the decays whose quark-level process is $b \rightarrow d e^\pm \mu^\mp$,
we assume ${\cal B}(\bar B^0 \rightarrow \pi^0 e^\pm \mu^\mp) = 0.5
{\cal B}(B^- \rightarrow \pi^- e^\pm \mu^\mp)$, and similarly for the
$\rho^0$, $\rho^-$ pair.  Again, we impose those constraints in the maximum
likelihood procedure, using information from both $\pi^-$ and $\pi^0$ modes but
quoting the ``average'' branching fraction
${\cal B}(B \rightarrow \pi e^\pm \mu^\mp) \equiv 0.5(
{\cal B}(B^- \rightarrow \pi^- e^\pm \mu^\mp)
+2 {\cal B}(\bar B^0 \rightarrow \pi^0 e^\pm \mu^\mp))$, and similarly with
the $\rho^0$, $\rho^-$ pair.  In all cases, by
${\cal B}( B \rightarrow h e^\pm \mu^\mp)$ we mean the {\it sum}
${\cal B}( B \rightarrow h e^+ \mu^-) +
{\cal B}( B \rightarrow h e^- \mu^+)$.

    Our search is thus for four different lepton-flavor-violating final states:
$e^\pm \mu^\mp$, $e^+ e^+$, $e^+ \mu^+$, and $\mu^+ \mu^+$; with four different
hadronic final states: $K$, $K^*$, $\pi$, $\rho$; a total of 16 decays.  For
each of the 16 decays, we maximize the likelihood ${\cal L}$, as a function of
signal branching fraction, by varying the yields of the two backgrounds.  (In so
doing, we constrain both backgrounds to be non-negative.)  The central value
obtained for signal is that giving the largest likelihood.  The statistical
significance of the signal is the square root of the difference in
$2 \ln {\cal L}$ between the maximum ${\cal L}$ and the ${\cal L}$ with signal
branching fraction set to zero.  If the largest likelihood corresponds to a
negative signal, we assign a significance of zero.
We find no compelling evidence for any of the decays.  All but
$B \rightarrow K^* e^\pm \mu^\mp$ have a statistical significance of less than
1.2 standard deviations, while $B \rightarrow K^* e^\pm \mu^\mp$ has a
statistical significance of 2.0 standard deviations.  In 16 searches, the
probability that one of the 16 will fluctuate up by at least 2 standard 
deviations is
$\sim$1/3, so our result is consistent with all branching fractions being zero, 
and no claim for a signal is being made.

    We obtain 90\% confidence level upper limits on the 16 branching fractions
by integrating the likelihoods, as a function of the assumed branching fraction,
from zero to that value which gives 90\% of the integral from zero to infinity. 
We increase the upper limit so found by 1.28 times the estimated systematic
error, which includes contributions from uncertainty in efficiency for detecting
the signal and uncertainty in the PDFs.  The upper limits are increased by
typically 12\% from these systematic error considerations.  Results are given in
Table~\ref{tab:limits}.  The limits on
decays to $\pi$, $K$ range from 1.0 to $2.0 \times 10^{-6}$, while those on
decays to $\rho$, $K^*$ range from 2.6 to $8.3 \times 10^{-6}$.

    As a check on the correctness of our continuum background PDFs, obtained
from Monte Carlo, we have analyzed the off-resonance data, both alone and with 
4 randomly chosen signal Monte Carlo events added.  We found no evidence of
`signal' in the off-resonance data, and the correct amount of signal (average
of 4.25, in 100 `toy experiments' for each of the 16 modes) when 
Monte Carlo signal events were added.

\begin{figure}
\begin{center}
\epsfxsize=7in
\epsfbox{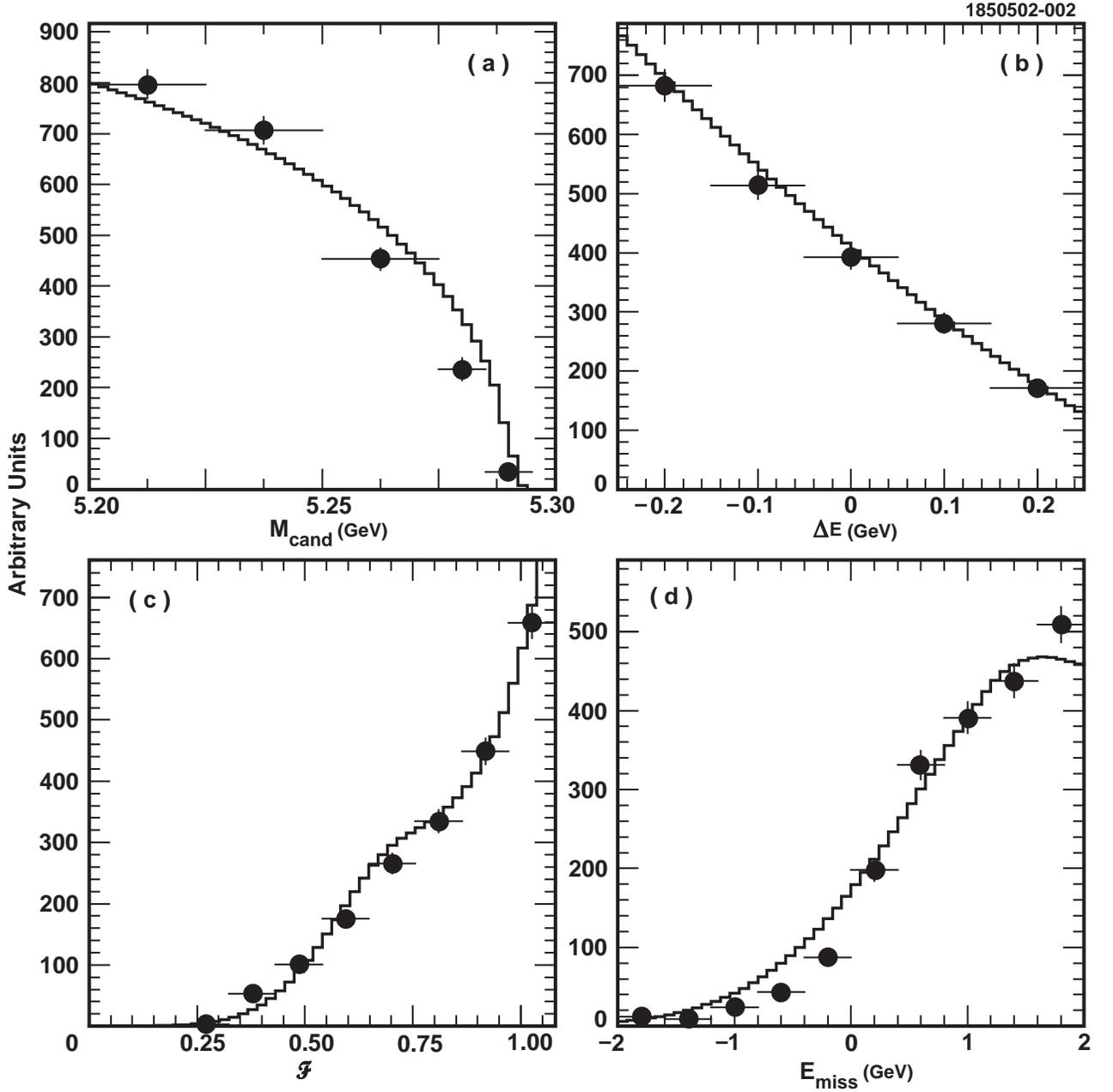}
\hfill \caption{
Results of the fit to the on-resonance data for the sum of the 16 modes 
with no signal allowed and the continuum background constrained to 
the scaled off-resonance yield. Distributions in (a) $M_{\rm cand}$, 
(b) $\Delta E$, (c) ${\cal F}$, and (d) $E_{\rm miss}$. Points are 
on-resonance data; solid histogram is the fit.
\label{fig:Fits}}
\end{center}
\end{figure}

    We have performed two checks on the correctness of our $B \bar B$
background PDFs. In the first, we added 4 randomly chosen signal
Monte Carlo events to the on-resonance data, and reanalyzed the data,
performing 100 such `toy experiments' on each of the 16 decay modes.
We found an average of 4.0 signal events, in agreement with the 
number added. This check shows that whatever bias is present in our 
analysis approximately cancels whatever real signal is present, an unlikely
coincidence unless both are small.
In the second check, we summed the on-resonance data sample
for the 16 decay modes, and fitted it, with no signal allowed in the fit and
with the continuum background constrained to the scaled off-resonance yield.
In Fig.~\ref{fig:Fits} we show the results of the fit for the distributions in
$M_{\rm cand}$, $\Delta E$, ${\cal F}$, and $E_{\rm miss}$. Agreement is
good. If instead we allowed signal in the fit and left the continuum 
background unconstrained (as in our actual analysis), we found 
4.0$^{+5.3}_{-4.0}$ signal events for 
the sum over 16 modes. From these checks we conclude that any 
bias is small, $\lesssim\frac{1}{2}$ event per mode, and is covered 
by our systematic error.


    In summary, we have searched for sixteen different lepton-flavor-violating
decays of the form $B \rightarrow h \ell \ell$.  We find no evidence for
any such decay, and place 90\% confidence level upper limits on the branching
fractions that range from 1.0 to $8.3 \times 10^{-6}$.  BaBar has limits on two
of these decays\cite{BaBar}, a factor of two more restrictive than ours.

We gratefully acknowledge the effort of the CESR staff in providing us with
excellent luminosity and running conditions.
This work was supported by 
the National Science Foundation,
the U.S. Department of Energy,
the Research Corporation, and
the Texas Advanced Research Program.

\begin{table}[ht]
\begin{center}
\begin{tabular}{|c|c|c|}
\hline
 Decay  mode &  Significance & Upper Limit  \\
             &   of Signal   & $(10^{-6})$  \\ \hline
$B \rightarrow K e^\pm \mu^\mp$      & 0.0$\sigma$   & 1.6     \\
$\ \ \ \ \ \ \ K^* e^\pm \mu^\mp$    & 2.0$\sigma$   & 6.2     \\
$\ \ \ \ \ \ \ \pi e^\pm \mu^\mp$    & 0.0$\sigma$   & 1.6     \\
$\ \ \ \ \ \ \  \rho e^\pm \mu^\mp$  & 0.6$\sigma$   & 3.2     \\ \hline
$B^+ \rightarrow K^- e^+ e^+$        & 0.0$\sigma$   & 1.0     \\
$\ \ \ \ \ \ \ \ K^{*-} e^+ e^+$     & 0.0$\sigma$   & 2.8     \\
$\ \ \ \ \ \ \ \ \pi^- e^+ e^+$      & 0.0$\sigma$   & 1.6     \\
$\ \ \ \ \ \ \ \ \rho^- e^+ e^+$     & 1.1$\sigma$   & 2.6     \\ \hline
$B^+ \rightarrow K^- e^+ \mu^+$      & 0.0$\sigma$   & 2.0     \\
$\ \ \ \ \ \ \ \ K^{*-} e^+ \mu^+$   & 0.0$\sigma$   & 4.4     \\
$\ \ \ \ \ \ \ \ \pi^- e^+ \mu^+$    & 0.0$\sigma$   & 1.3     \\
$\ \ \ \ \ \ \ \ \rho^- e^+ \mu^+$   & 0.3$\sigma$   & 3.3     \\ \hline
$B^+ \rightarrow K^- \mu^+ \mu^+$    & 0.0$\sigma$   & 1.8     \\
$\ \ \ \ \ \ \ \ K^{*-} \mu^+ \mu^+$ & 0.5$\sigma$   & 8.3     \\
$\ \ \ \ \ \ \ \ \pi^- \mu^+ \mu^+$  & 0.0$\sigma$   & 1.4     \\
$\ \ \ \ \ \ \ \ \rho^- \mu^+ \mu^+$ & 1.0$\sigma$   & 5.0     \\ \hline
\end{tabular}
\end{center}
\caption{
For each of 16 decay modes, the statistical significance of the signal,
and the 90\% confidence level upper limit on the branching fraction, including
systematic error.  In the modes $B \rightarrow h e^\pm \mu^\mp$, the limit
quoted is on the {\it sum} ${\cal B}(B \rightarrow h e^+ \mu^-) +
{\cal B}(B \rightarrow h e^- \mu^+)$.
\label{tab:limits}}
\end{table}

\end{document}